\documentstyle[aps,preprint]{revtex}
\input{epsf}
\begin{document}
\setlength{\leftmargin}{5em}
\setlength{\rightmargin}{0em}
\newpage
\baselineskip 16pt plus 2pt minus 2pt
\par
\topmargin=-2cm      % distance from top of the page to first line
                     % of text plus one inch
\vspace*{.5in}
%\hfill{DUKE-TH-95-88}

%\hfill{DOE/ER/40427-06-N95}
%\vspace{10.0pt}

\begin{centering}
{\Large\bf Neutral Pion Photoproduction on Nuclei in
Baryon Chiral Perturbation Theory}\\

\vspace{40.0pt}
{{\bf S.R.~Beane}$^{1,}$\footnote{E-mail address: sbeane@phy.duke.edu},
{\bf C.Y.~Lee}$^{2,}$\footnote{E-mail address: clee@utpapa.ph.utexas.edu}
and {\bf U.~van Kolck}$^{3,}$\footnote{E-mail address: vankolck@alpher.npl.
washington.edu} }\\
\vspace{20.0pt}
{\sl $^{1}$Department of Physics,
Duke University, Durham, NC 27708} \\
\vspace{15.0pt}
{\sl $^{2}$Nuclear Theory Group} \\
{\sl Department of Physics,
The University of Texas, Austin, TX 78712} \\
\vspace{15.0pt}
{\sl $^{3}$Department of Physics, FM-15,
University of Washington, Seattle, WA 98195} \\
\end{centering}
\vspace{20.0pt}
\begin{center}
\begin{abstract}
Threshold neutral pion photoproduction on light nuclei is studied in
the framework of baryon chiral perturbation theory. We obtain a
general formula for the electric dipole amplitude in the special case
of neutral pion photoproduction on a nucleus. To third order in small
momenta, the amplitude is a sum of 2- and 3-body interactions with no
undetermined parameters.  With reasonable input from the single
nucleon sector, our result for neutral pion photoproduction on the
deuteron is in agreement with experiment.
\end{abstract}

\vspace*{10pt}
%\noindent

\vfill
\end{center}
\newpage
%Body
%%%%%%%%%%%%%%%%%%%%%%%%%%%%%%%%%%%%%%%%%%%%%%%%%%%%%%%%%%%%%%%%%%%%%%%
\section{Introduction}

In recent times, there has been increasing interest in applying the
method of chiral lagrangians, or chiral perturbation theory (${\chi PT}$),
to processes involving more than a single
nucleon [1-8].
This interest is motivated by the desire to determine what aspects of
nuclear physics can be understood on the basis of the chiral symmetry
of $QCD$. In the limit of vanishing $u$ and $d$ quark masses the $QCD$
lagrangian admits a global $SU(2)_L\times SU(2)_R$ chiral symmetry.
The absence of parity doubling in the hadronic spectrum implies that
this chiral symmetry is either anomalous, or spontaneously broken by
the $QCD$ vacuum. The existence of suitable Goldstone boson
candidates, the pions, bears out the latter conjecture.  {\it A
priori}, the space of chiral symmetric operators involving the
relevant degrees of freedom --- nucleons and pions--- is infinite.
Moreover, since the interactions are strong, nothing is really learned
insofar as $QCD$ is concerned by restricting oneself to any arbitrary
finite subset of chiral symmetric operators; e.g., the simplest
operators involving the fewest number of fields, or renormalizable
operators.  Fortunately, as a consequence of non-linearly realized
chiral symmetry, low-energy hadronic matrix elements involving pions
are analytic in momenta.  The parameters that appear at leading order
in small momenta are well known, and so chiral symmetry leads to
non-trivial predictions at energies near the threshold of a physical
process --- so-called low-energy theorems (LETs).  The crucial fact
that the $u$ and $d$ quark masses are not identically zero can be
straightforwardly accounted for in perturbation theory.  As with any
approximation scheme, the fundamental importance of ${\chi PT}$ is its
ability to handle corrections to the leading order in a systematic
way\cite{physca}.  This method has been applied with great success to
the interactions of pions with a single nucleon\cite{ulf}\cite{ulf2}.
One might then wonder what generic features of nuclear physics can be
deduced from this chiral symmetry of $QCD$.

The main technical difficulty that arises when considering more than a
single nucleon is that ${\chi PT}$ necessarily breaks down, as is made clear
by the
appearance of shallow nuclear bound states\cite{swnp2}.  This
breakdown manifests itself via infrared singularities in Feynman
diagrams evaluated in the static approximation.  The problem is clear
in the language of time-ordered perturbation theory.  Evidently there
are two types of energy denominators that can appear in a typical
time-ordered graph.  The first type arises from intermediate states
which differ in energy from the initial and final states only by the
emission or absorption of soft pions. These energy denominators are of
the order of a small momentum or pion mass and are therefore
consistent with the usual chiral power counting scheme.  On the other
hand, the second type of intermediate states differ in energy from the
initial and final states by a nucleon 3-momentum, and therefore blow
up in the static limit.  Graphs of the first type are called
irreducible.
Following the tenets of
scattering theory, one can modify the rules and use ${\chi PT}$ to
calculate an effective potential, which consists of the sum of all
$N_n$-nucleon irreducible graphs\cite{swnp1}\cite{swnp2}.  The
S-matrix, which of course includes all reducible contributions, is then
obtained
through iteration by solving a Lippmann-Schwinger equation.  Several
generic features of nuclear physics, including two- and few-body
forces \cite{swnp2}\cite{ubi1}\cite{ubi2} and isospin violation \cite{ubi3},
have been shown to arise naturally in this approach.

Here we will apply chiral perturbation theory to a scattering process
involving light nuclei. This program was initiated several years ago
by Weinberg\cite{swnp3}, in a study of the pion-nucleus scattering
lengths.  The strategy is best described graphically. In figure 1 we
display the anatomy of a scattering matrix element. ${\chi PT}$ generates
the irreducible kernel, which is then sewn to the external nuclear
wave functions (with $N_n=A$).
Nuclear wave functions, too,
are calculable in ${\chi PT}$, and in fact exist for the
deuteron\cite{ubi1}. A completely consistent calculation can
be carried out, and results are reported here.
However, this
approach is in a
sense counterproductive, since it is then unclear what aspect of
chiral symmetry is being tested.
In the spirit of ${\chi PT}$,
where experimental input is always welcome, one should make use of the
most successful phenomenological potential. This allows one to test
the relevance of chiral symmetry in determining the irreducible
scattering kernel.  In the case of the deuteron, we also use the well known
Bonn potential wave function\cite{bonn}, which although not respectful of
chiral
symmetry, is at least spiritually linked to $QCD$.

In this paper, we consider pion photoproduction on light nuclei at
threshold. We carry out a general power counting analysis of
photoproduction, but specialize to neutral pion photoproduction. We
calculate a general formula for the invariant threshold amplitude to
third order in small momenta.  This process is of course interesting
in its own right; there has been no systematic calculation based on
chiral symmetry. Furthermore, as we will see, this process is
intimately related to threshold neutral pion photoproduction on a
single nucleon, a process which has caused a great deal of theoretical
and experimental confusion.  In contrast to pion-nucleus scattering at
threshold\cite{swnp3}, the amplitude for neutral pion photoproduction
on nuclei, to third order in small momenta, has no undetermined
parameters. Unfortunately, although the same is true in the single
nucleon sector, evidently the amplitude there converges slowly at
best. Hence the single-scattering contribution must be treated as
phenomenological input. With a reasonable choice for this
contribution, our result for the deuteron electric dipole amplitude is
in agreement with experiment.

This paper is organized as follows. In section 2 we review the
standard power counting formulas. In section 3 we present the heavy
fermion effective lagrangian to the order relevant to our calculation.
Section 4 consists of a power counting analysis of pion
photoproduction on a nucleus composed of $A$ nucleons. Here we give a
general formula for the neutral pion photoproduction amplitude . We
specialize to neutral pion photoproduction on the deuteron in section
5. Finally, we summarize and conclude. Several appendices are included
for pedagogical purpose.

%%%%%%%%%%%%%%%%%%%%%%%%%%%%%%%%%%%%%%%%%%%%%%%%%%%%%%%%%%%%%%%%%%%%
\section{Power Counting}

In the absence of exact solutions to quantum mechanical equations of
motion, as in $QCD$, systematic statements are possible only when a
small dimensionless expansion parameter is identified.  One class of
dimensionless parameters consists of pure numbers, such as a coupling
constant associated with a renormalizable interaction, or the inverse
of the dimensionality of a group, as in the large-${\cal N}_{c}$
limit.  A second class of dimensionless parameters consist of
ratios of dimensional quantities. This class arises naturally via
broken symmetries, and is the case of interest in this paper.

Spontaneously broken continuous symmetries give rise to massless
Goldstone modes. These modes do not propagate in the vacuum, and
therefore only couple derivatively. Hence at energies small relative
to the characteristic symmetry breaking scale, the interactions of
Goldstone bosons admit a power series in momenta.  In $QCD$ the small
parameter is $Q/{\Lambda _\chi}$, where $Q$ is a characteristic
momentum, and ${\Lambda _\chi}$ is the scale of chiral symmetry
breaking ---of order the masses of the lowest lying resonances. The
effects of non-zero quark masses give the pion a mass. Since
${M_\pi}/{\Lambda _\chi}$ is small, we treat $Q$ as representing a
small momentum or a pion mass.  A generic matrix element involving the
interaction of any number of pions and nucleons can then be written in
the form

\begin{equation}
{\cal M}={Q^\nu}{\cal F}(Q/\mu),
\end{equation}
where $\mu$ is a renormalization scale, and $\nu$ is a counting
index.  It is straightforward to arrive at a general formula for $\nu$
by considering the momentum space structure of generic Feynman rules. In
this
way one finds\cite{swnp2}

\begin{equation}
\nu =4L-{I_n}-2{I_\pi}+{\sum _i}{V_i}{d_i}+4-4C,
\end{equation}
where $L$ is the number of loops, $I_{n,\pi}$ is the number of internal
fermion, pion
lines, $V_i$ is the number of vertices of type $i$, $d_i$ is
the number of derivatives or powers of $M_\pi$ which contribute to an
interaction of type $i$, and $C$ is the number of
separately connected pieces. One can also make use of the graph theoretic
identities:

\begin{eqnarray}
&L&={I_n}+{I_\pi}-{\sum _i}{V_i}+1 \\
& &2{I_n}+{E_n}={\sum _i}{V_i}{n_i},
\end{eqnarray}
where $E_n$ is the number of external nucleon lines, and $n_i$ is the
number of fermion
fields involved in an interaction of type $i$.
Here we are interested in processes with the same number of nucleon lines
in the initial and final state, and so defining
${N_n}\equiv{E_n}/2$, we obtain the master formula

\begin{eqnarray}
\nu &=&4-{N_n}-2C+2L+{\sum _i}{V_i}{\Delta _i}\nonumber \\
& &{\Delta _i}\equiv {d_i}+{n_i}/2-2.
\end{eqnarray}
This formula is important because chiral symmetry places a lower
bound: ${\Delta _i}\geq 0$. Hence the leading {\it irreducible} graphs
are tree graphs ($L=0$) with the maximum number $C$ of seperately
connected pieces, constructed from vertices with $\Delta _i =0$.

How is this analysis altered in the presence of an
electromagnetic field? Photons couple via the electromagnetic
field strength tensor and by minimal substitution.
This has the simple effect of modifying the lower bound on $\Delta _i$ to
${\Delta _i}\geq -1$. (And, of course, of introducing an expansion in
the electromagnetic coupling.)

%%%%%%%%%%%%%%%%%%%%%%%%%%%%%%%%%%%%%%%%%%%%%%%%%%%%%%%%%%%%%%%%%%%%%%%
\section{Baryon ${\chi PT}$}

With the power counting scheme established, the next step is to construct
the various interactions which contribute to matrix elements for a given
value of $\nu$.
The technology that goes into building an effective lagrangian
is standard by now\cite{ulf}. Here we establish our conventions.  The pion
triplet is contained in a matrix field

\begin{equation}
\Sigma =\exp ({\frac{i{\vec\pi}\cdot{\vec\tau}}{f_\pi}}),
\end{equation}
which transforms under $SU(2)_L\times SU(2)_R$ as
$\Sigma\rightarrow L\Sigma{R^\dagger}$. It is convenient to introduce the
field $\xi\equiv\sqrt\Sigma$, with transformation property
$\xi\rightarrow L\xi{U^\dagger}=U\xi{R^\dagger}$. This transformation
property implicitly defines $U$. Out of $\xi$ one can construct

\begin{eqnarray}
{V_\mu}&=&\frac{1}{2}[{\xi ^\dagger}({\partial _\mu}-ie{{\cal A}_\mu}Q)\xi +
 \xi({\partial _\mu}-ie{{\cal A}_\mu}Q){\xi ^\dagger}] \\
{A_\mu}&=&\frac{i}{2}[{\xi ^\dagger}({\partial _\mu}-ie{{\cal A}_\mu}Q)\xi -
 \xi({\partial _\mu}-ie{{\cal A}_\mu}Q){\xi ^\dagger}],
\end{eqnarray}
which transform as ${V_\mu}\rightarrow U{V_\mu}{U^\dagger} +
U{\partial _\mu}{U^\dagger}$ and ${A_\mu}\rightarrow
U{A_\mu}{U^\dagger}$ under $SU(2)_L\times SU(2)_R$.
It is convenient to assign the
nucleon doublet $N$ the transformation property $N\rightarrow UN$.
With these ingredients, one can construct the leading order
effective lagrangians,

\begin{eqnarray}
{{\cal L}^{(2)}_{\pi\pi}}&=&\frac{1}{4}{f_\pi^2}tr({D_\mu}{\Sigma ^\dagger}
    {D^\mu}{\Sigma})
    +\frac{1}{4}{{f_\pi^2}{M_\pi^2}}tr({\Sigma}+{\Sigma ^\dagger})\\
{{\cal L}^{(1)}_{\pi N}}&=&i{\bar {N}}({{D}\!\!\!\!/}-m){N}
               + {g_A}{\bar {N}}{A\!\!\!/}{\gamma _5}{N}\\
{{\cal L}^{(0)}_{N N}}&=&\frac{1}{2}{C_a}({\bar N}{\Gamma _a}{N})^2,
\end{eqnarray}
where ${D_\mu}\equiv{\partial _\mu}+{V_\mu}$, $\Gamma _a$ is an
arbitrary Hermitian operator, and the $C_a$ are undetermined
coefficients. The pion covariant derivative is
${D_\mu}\Sigma={\partial _\mu}\Sigma -ie{{\cal A}_\mu}[Q,\Sigma ]$, where
$Q=(1+{\tau _3})/2$. The appearance of the nucleon mass $m$, both
explicitly and through the time derivative acting on the nucleon
field, implies the existence of a dimensionless quantity that is not
small: $m/{\Lambda _\chi}
\sim 1$. This destroys the power counting. Fortunately, since the
nucleon carries a quantum number ---Baryon number--- which is
conserved by the strong interactions, one can maintain a consistent
power counting framework by choosing a heavy-fermion basis in which
the nucleon mass does not appear at leading order\cite{hg}\cite{jm}.
The nucleon momentum can be written as ${p^\mu}=m{v^\mu}+{k^\mu}$,
where ${v^\mu}$ is the nucleon four-velocity (${v^2}=1$), and
${k^\mu}$ is the amount by which the nucleon momentum is off shell. We
can then define a velocity dependent basis

\begin{eqnarray}
{B_v}(x)={e^{im{v\cdot x}}}N(x).
\end{eqnarray}
In the rest frame of the nucleon, ${v^\mu}=(1,0,0,0)$.
Although choosing a particular velocity breaks covariance, integrating over
all velocity-dependent lagrangians restores covariance\cite{hg}.
The velocity-dependent lagrangian in the new basis is

\begin{equation}
{\cal L}_v^{(1)}=i{\bar {B_v}}(v\cdot D){B_v}
               + 2{g_A}{\bar {B_v}}(A\cdot {S_v}){B_v} ,
\end{equation}
where $S_v^\mu$ is the spin operator.
Note that the nucleon mass term is no longer present in the lagrangian.
The new effective expansion parameter is ${|k|}/{\Lambda _\chi}$.  The $1/{m}$
corrections enter via higher dimensional operators which affect the
higher orders in ${\chi PT}$. At next to leading order one finds,

\begin{eqnarray}
{\cal L}_v^{(2)}&=&\frac{1}{2m}{\bar {B_v}}({-D^2}+(v\cdot D)^2
                    +2i{g_A}\{{v\cdot A},{S\cdot D}\} \nonumber \\
& &-\frac{i}{2}[{S_v^\mu},{S_v^\nu}][(1+{\kappa _v}){f_{\mu\nu}^+}+
\frac{1}{2}({\kappa _s}-{\kappa _v})tr{f_{\mu\nu}^+}]+...){B_v},
\end{eqnarray}
where  ${\kappa _v}={\kappa _p}-{\kappa _n}$, ${\kappa _s}={\kappa _p}+
{\kappa _n}$, and
${f_{\mu\nu}^+}\equiv e({\xi ^\dagger}Q\xi + \xi Q{\xi ^\dagger})
{F_{\mu\nu}}$.  ${F_{\mu\nu}}$ is the electromagnetic
field strength tensor. The dots signify that we have included only the
interactions at this order which are relevant to our calculation.
Note, finally, that the $\Delta$ isobar can be introduced in an analogous
way.

%%%%%%%%%%%%%%%%%%%%%%%%%%%%%%%%%%%%%%%%%%%%%%%%%%%%%%%%%%%%%%%%%%%%%%%
\section{Pion photoproduction on light nuclei}

Now we can put our technology to use.  Consider a process with $N_n=A$
nucleons in both the initial and the final state, and a single photon
and a single pion in the initial and final state,
respectively. Limiting ourselves to lowest order in the
electromagnetic coupling, we can order the chiral expansion of the
irreducible diagrams by way of the counting index $\nu$:

\begin{center}
\underline{${\bf \nu =3-3A}$}
\end{center}
\noindent At leading order in small momenta, the matrix element is given by
tree graphs with the maximum number of separately connected pieces
($L=0$, $C=A$) constructed out of
one interaction with $\Delta _i =-1$, and interactions with $\Delta _i =0$.
For example, at this order one has the the Kroll-Ruderman term (figure 2a).

\begin{center}
\underline{${\bf \nu =4-3A}$}
\end{center}
\noindent The first corrections to the leading terms are still
tree graphs with the maximum number of separately connected pieces
($L=0$, $C=A$), but have one vertex of higher index:
i) either one vertex with $\Delta _i =-1$ and one with $\Delta _i =1$;
ii) or vertices with $\Delta _i =0$, the one involving the photon field
being a $1/m$ correction (figure 2b).

\begin{center}
\underline{${\bf \nu =5-3A}$}
\end{center}
\noindent There are four classes of corrections at this order:

(i) One loop graphs $(L=1)$ with interactions with $\Delta _i =-1,0$,
and $C=A$ (figure 2c).

(ii) Counterterm graphs with $L=0$, $\Delta _i =1$ and $C=A$ (figure 2d).
In the case of neutral pion photoproduction, these graphs vanish and
only finite loop graphs remain.

(iii)
$C=A$ tree ($L=0$) graphs with i) $\Delta _i =-1,2$ interactions;
or ii) $\Delta _i =0,1$ interactions (figure 2e)
\footnote{ Note that there are no irreducible graphs of this type
where the photon and the pion are attached to different nucleons, because
energy and momentum conservation require
an interaction between nucleons (for example, by pion exchange),
%the exchange of a pion
and somewhere along the nucleon lines there is an energy flow
of the order of the pion mass.
The irreducible part of the diagram then contains the
nucleon interaction, which decreases the number of separately connected
pieces and renders the diagram higher order. At higher energies,
where higher order contributions might become more important,
it may well be necessary to account for such effects \cite{ktb}.}
(Some of these are $1/{m^{2}}$ corrections and are proportional to
the nucleon magnetic moments.)

(iv) Finally, there are tree graphs $(L=0)$ with one less than the
maximum number of separately connected pieces $(C=A-1)$, and interactions with
$\Delta _i =-1,0$. These graphs fall into two separate classes. There are
the 3-body graphs like the Feynman graph of figures 2f and the time-ordered
graph of 2g. For $A>2$, there are also disconnected 2-body interactions as in
figure 2h.

In principle, all of these graphs contribute
to pion photoproduction on nuclei. However, some generic simplifications
arise when one sews in the nuclear wavefunctions.

First, the time-ordered graphs of type 2g and 2h get cancelled
against recoil in the one-pion-exchange piece of the potential.
In order to see this, consider the three diagrams of figure 3.
These graphs all have the same spin-isospin structure: they differ
only in the energy denominators. The first two graphs (figures 3a and 3b)
arise when a diagram
like 2a is sandwiched between wavefunctions obtained from a potential
whose long range part comes from pion exchange. They are proportional
to

\begin{eqnarray*}
\frac{1}{E(\vec{p}_{1}-\vec{q})+E(\vec{p}_{2}+\vec{q})-E(\vec{p}_{1})
 -E(\vec{p}_{2})} \makebox[1.0in][l]{$\times$} &  &   \\
  \times\left[\frac{1}{E(\vec{p}_{1}-\vec{q})+w-E(\vec{p}_{1})}
  +\frac{1}{E(\vec{p}_{2}+\vec{q})+w-E(\vec{p}_{2})}\right] & = &
\end{eqnarray*}
\begin{eqnarray}
& = & \frac{2}{w[E(\vec{p}_{1}-\vec{q})+E(\vec{p}_{2}+\vec{q})
      -E(\vec{p}_{1})-E(\vec{p}_{2})]}  \nonumber  \\
&   & \makebox[2.0in][r]{$-$} \frac{1}{w^{2}}\left(1+O\left(
      \frac{E}{w}\right)\right).
\end{eqnarray}
The first term corresponds to static one-pion-exchange in the potential;
it is big, as antecipated, because these reducible diagrams have small
nucleon energy denominators. The second term is smaller because of the
additional small recoil numerator, while the dots sum higher orders in
chiral perturbation theory. On the other hand, the remaining graph
(figure 3c) is proportional to

\begin{equation}
  \frac{1}{w^{2}} \left( 1+O \left(\frac{E}{w}\right) \right),
\end{equation}
and exactly cancels recoil in the reducible diagrams. Similar cancellations
can be found
among the other time ordered diagrams. In other words, to this order in chiral
perturbation theory we can omit the time ordered diagrams 2g and 2h by
at the same time disregarding the energy dependence in the potential.

Second, in the case of neutral pion photoproduction at threshold
a number of the graphs in figure 2 will not contribute: those
where the photon line is attached to a pion, and those that go
like $S \cdot q$, where $q$ is the outgoing pion momentum.
In particular,
all the leading order graphs vanish (figure 2a), which immediately
suggests that the cross section will be smaller than for charged
pion production, and --- of particular interest to us --- more sensitive
to two-nucleon contributions.
Moreover,
the 3-body time-ordered
graphs (figure 2g) and the two body disconnected graphs (figure 2h)
also vanish, so we can expect little influence of the energy-dependent
part of the potential. As noted above, the loop graphs are finite for
a neutral pion, and so there are no undetermined parameters to this order.

Since all single-scattering contributions have been calculated to third
order in small momenta, all that is left to calculate are the 3-body
graphs of figure 4. In Coulomb gauge ($v \cdot \epsilon=0$),
only figures 4a and 4b survive.
Hence, to $\vartheta (q^3)$ in ${\chi PT}$, the threshold amplitude for
neutral pion photoproduction on a nucleus is remarkably simple.
We obtain the general formula

\begin{equation}
{{\cal M}_{\Psi _A}}{|_{q=0}}={{\cal M}_{\Psi _A}^{ss}}+
{{\cal M}_{\Psi _A}^{(a)}}+{{\cal M}_{\Psi _A}^{(b)}},
\end{equation}
where

\begin{eqnarray}
{{\cal M}_{\Psi _A}^{ss}}&=&{S_{\Psi _A}}(k) {\sum _{i}}{{\cal M}^i_N} \\
{{\cal M}_{\Psi _A}^{(a)}} & =&  i\frac{2eg_{A}M_{\pi}m}{(2\pi )^{3}
f_{\pi}^{3}}
{\sum _{i<j}}<{\Psi _A}|({\vec\tau}^{(i)}\cdot {\vec\tau}^{(j)}
-\tau^{(i)}_{z}\tau^{(j)}_{z})
\frac{(\vec{J}\cdot\vec\epsilon )}{\vec{{q_{ij}}}^{\prime^{2}}}|{\Psi _A}>  \\
{{\cal M}_{\Psi _A}^{(b)}} & =&  i\frac{4eg_{A}M_{\pi}m}{(2\pi )^{3}
f_{\pi}^{3}}
{\sum _{i<j}} <{\Psi _A}|({\vec\tau}^{(i)}\cdot {\vec\tau}^{(j)}
-\tau^{(i)}_{z}\tau^{(j)}_{z})
\frac{ [\vec{J}\cdot({{\vec{{q_{ij}}}^{\prime}} -\vec{k}})]
(\vec{{q_{ij}}}^{\prime}\cdot\vec\epsilon)}
{[({\vec{{q_{ij}}}^{\prime}} -\vec{k})^2
+M_{\pi}^{2}] \vec{{q_{ij}}}^{\prime^{2}}} |{\Psi _A}>.
\end{eqnarray}
${S_{\Psi _A}}(k)$ is a generic overlap function.
(See next section and appendices for details.)

The single-scattering electric dipole amplitudes have been calculated
to ${{\vartheta (q^3)}}$ without an explicit isobar field, and are
given by\cite{ber1}

\begin{eqnarray}
{E_{0+}^{{\pi ^0}p}}&=&-\frac{e{g_A}}{8\pi{f_\pi}}\{ \frac{M_\pi}{m}-
\frac{M_\pi ^2}{2m^2}(3+{\kappa _p})-\frac{M_\pi ^2}{16{f_\pi ^2}}\}\nonumber\\
{E_{0+}^{{\pi ^0}n}}&=&-\frac{e{g_A}}{8\pi{f_\pi}}\{
\frac{M_\pi ^2}{2m^2}{\kappa _n}-\frac{M_\pi ^2}{16{f_\pi ^2}}\}.
\end{eqnarray}
Unfortunately, the single nucleon sector is not well understood.  On
one hand, the neutron amplitude has not been measured. On the other
hand, the electric dipole amplitude, ${{ E_{0+}^{{\pi_0}p}}}$, has an
interesting ---and rather complicated--- history, which we will
briefly discuss here
\footnote{For a review, see \cite{tia} .}.
An expansion of the amplitude in powers of the pion energy was first used
to derive a tree level ``LET'' \cite{vain}
%A tree level ``LET'' was first derived in \cite{vain}
yielding a value ${{ E_{0+}^{{\pi_0}p}}}
=-2.23\cdot{10^{-3}}/{{M_{\pi^+}}}$; this LET was later rederived
\cite{nkf} under an extended PCAC hypothesis, in a way that made
explicit that it also included loop corrections, like rescattering.
Experiments at Mainz\cite{mainz} and Saclay\cite{saclay} suggested a
violation of this LET.  Subsequently, the data were reexamined,
leading to the revised value ${{ E_{0+}^{{\pi_0}p}}} =(-2.0\pm
0.2)\cdot{10^{-3}}/{{M_{\pi^+}}}$, a result in agreement with the
``classical'' LET\cite{bh}. The source of the discrepancy is isospin
violation; the difference of $6.8\;MeV$ between the $p{\pi^0}$ and
$n{\pi^+}$ threshold leads to a rapid variation of the amplitude in
this region.  The correct interpretation of the data depends
critically on the details of this variation.  The situation as of 1991
is reviewed in \cite{bh}.  To further complicate the matter, it was
then found that there are additional {\it large} finite loop
contributions to ${{ E_{0+}^{{\pi_0}p}}}$ at ${{\vartheta (q^3)}}$;
the so-called triangle graphs (see figure
2c)\cite{ber1}\cite{ber1a}\cite{ber2}.  They are of higher order in
the expansion in pion energy, but are proportional to $M_{\pi}^{-1}$
\cite{ber1}\cite{skf}.  With these contributions ---which clearly must
be included--- the ${{\vartheta (q^3)}}$ LET no longer seems to agree
with the data. Moreover, it is clear that ${{ E_{0+}^{{\pi_0}p}}}$ is
---at best--- slowly converging.  ${{ E_{0+}^{{\pi_0}p}}}$ has now
been calculated to ${{\vartheta (q^4)}}$, and shows no signs of
converging.  Evidently, it is difficult to escape the conclusion that
the s-wave multipole is not a good testing ground of QCD. All is not
lost however; recently novel p-wave LET's have been calculated, and
found to have better convergence properties than the
s-waves\cite{ber3}. Nevertheless, this failure of ${\chi PT}$ in
describing the s-waves without explicit isobar fields is a lesson that
cannot be ignored here.  Evidently the sensible thing to do is to make
a best phenomenological estimate of the single-scattering
contribution.  The chiral prediction without the triangle graphs would
appear to be a reasonable phenomenological estimate of ${{
E_{0+}^{{\pi_0}p}}}$, and so we assume the same for ${{
E_{0+}^{{\pi_0}n}}}$.

%%%%%%%%%%%%%%%%%%%%%%%%%%%%%%%%%%%%%%%%%%%%%%%%%%%%%%%%%%%%%%%%%%%%%%%
\section{Neutral pion photoproduction on the deuteron}

Here we will consider the deuteron.  Deuteron phase shifts and
properties are well described by the Bonn wave function\cite{bonn}.
We also give ---in paranthesis--- the chiral wave function\cite{ubi1}
results.  With the conventions defined in the appendices, the
single-scattering contribution to the deuteron is given by

\begin{equation}
{E^{ss}_{d}}=
\frac{1+{M_\pi}/{m}} {1+{M_\pi}/{m_d}}
({E_{0+}^{{\pi ^0}p}}+{E_{0+}^{{\pi ^0}n}} ){{S_d}(k/2)}=
{-1.34\times {10^{-3}}}/{{{M_{\pi^+}}}}\qquad (-1.38),
\end{equation}
where ${S_d}(k/2)$ is the deuteron form-factor:

\begin{equation}
{S_d}(k/2)=\int d^{3}{p} {{\phi}}^{\ast}_{f}({p})\;
\;{{\phi}}_{i}({p}-\frac{k}{2})=0.722\qquad (0.742),
\end{equation}
evaluated with the Bonn and chiral wave functions, respectively, and
we have used the phenomenological estimates:

\begin{eqnarray}
{E_{0+}^{{\pi ^0}p}}&=&-\frac{e{g_A}}{8\pi{f_\pi}}\{ \frac{M_\pi}{m}-
\frac{M_\pi ^2}{2m^2}(3+{\kappa _p})\}
={-2.24\times {10^{-3}}}/{{{M_{\pi^+}}}}  \nonumber \\
{E_{0+}^{{\pi ^0}n}}&=&-\frac{e{g_A}}{8\pi{f_\pi}}\{
\frac{M_\pi ^2}{2m^2}{\kappa _n}\}={0.5\times {10^{-3}}}/{{{M_{\pi^+}}}} .
\end{eqnarray}
(The corresponding result from the ``complete'' ${{\vartheta (q^3)}}$
Eq.(21) will also be discussed shortly.)  The 3-body contributions are
readily obtained (see appendices for details).  Figure 3a yields

\begin{eqnarray}
{{E_d}^{(a)}}&=&
-\frac{eg_{A}m_{\pi}m}{{8\pi ({M_\pi}+{m_d})}\pi {f_{\pi}^{3}}k}
\int_{0}^{\infty}\;\frac{U^{2}}{r^{2}}\;\sin(\frac{kr}{2})\;dr\nonumber\\
&=&{-2.20\times {10^{-3}}}/{{{M_{\pi^+}}}}\qquad (-2.18)
\end{eqnarray}
and figure 3b yields

\begin{eqnarray}
{{E_d}^{(b)}}&=&
-\frac{eg_{A}m_{\pi}m}{{8\pi ({M_\pi}+{m_d})}2\pi {f_{\pi}^{3}}}
\int_{0}^{1}dz
\int_{0}^{\infty}dr\;e^{-m^{\prime}r}{U^2}(\frac{1}{r}
\frac{\sin{[(z-\frac{1}{2})kr]}}{(z-{\frac{1}{2}})kr}\nonumber\\
& &-(\frac{1}{r}+m^{\prime})\;
\{\frac{\sin(z-{\frac{1}{2}} )kr}{[(z-{\frac{1}{2}})kr]^{3}}\;-\;
\frac{\cos(z-{\frac{1}{2}})kr}{[(z-{\frac{1}{2}})kr]^{2}} \})\nonumber\\
&=&{-0.43\times {10^{-3}}}/{{{M_{\pi^+}}}}\qquad (-0.39).
\end{eqnarray}
The total is then given by

\begin{equation}
{E_d}={{E_d}^{ss}}+{{E_d}^{(a)}}+{{E_d}^{(b)}}
={-3.97\times {10^{-3}}}/{{{M_{\pi^+}}}}\qquad (-3.95),
\end{equation}
to be compared to the  experimental value\cite{arg}:

\begin{equation}
{{E_d}^{exp}}={(-3.74\pm 0.25)\times {10^{-3}}}/{{{M_{\pi^+}}}}.
\end{equation}
Hence the simple picture provided by chiral symmetry does fairly well.
In particular, the importance of the 3-body correction (charge
exchange contribution) emerges as a consequence of chiral symmetry.
It is gratifying that the results are not particularly sensitive to
the details of the wave function.  There are of course several serious
caveats.  Strictly speaking the ${{\vartheta (q^3)}}$ result without
explicit isobars fairs badly, as is made clear in Table 1. This is, of
course, a consequence of the theoretical failure in the single nucleon
sector.  Better experimental data is necessary in the single
scattering sector.  Currently new measurements of ${{
E_{0+}^{{\pi_0}p}}}$ are underway at Mainz and Saskatoon\cite{ulf2}.
In principle, with an accurate measurement of the deuteron
photoproduction amplitude one could extract the neutron electric
dipole amplitude using our results.  Of course, in order to be
convinced of the soundness of this method one would have to calculate
${{\vartheta (q^4)}}$ 3-body effects in order to test the convergence
properties of the nuclear matrix elements.  Finally, we note that our
final results are quite similar to results obtained some time ago,
based on the photoproduction low-energy theorems in the impulse
approximation, and assorted estimates of the three-body
corrections\cite{koch}\cite{laget}\cite{faldt}.  One might say that we
have placed these successful results on a more sound theoretical
footing by determining ---by way of chiral power counting--- the
precise graphs that should dominate at threshold.

%%%%%%%%%%%%%%%%%%%%%%%%%%%%%%%%%%%%%%%%%%%%%%%%%%%%%%%%%%%%%%%%%%%%%%%
\section{Conclusion}

We have taken a pedagogical approach to pion photoproduction on nuclei
in the framework of baryon chiral perturbation theory.  The method
presented allows one to make systematic use of chiral symmetry in a
scattering process involving nuclei.  In general, calculations are
more involved than those of the single nucleon sector, since one must
focus on the set of irreducible graphs, which requires use of
time-ordered perturbation theory.  In the special case of neutral pion
photoproduction, the amplitude to ${{\vartheta (q^3)}}$ is simple,
involving only tree level Feynman graphs evaluated in the
heavy-fermion formalism. We evaluated the deuteron electric dipole
amplitude using the Bonn and chiral wave functions, together with a
phenomenological estimate for the single nucleon (impulse
approximation) contributions.  The result of this calculation, like
that of the $\pi$-deuteron scattering length\cite{swnp3}, is in
agreement with experiment.  Hence the importance of 3-body
contributions emerges in both cases as a consequence of chiral
symmetry. The result is, of course, critically dependent on input from
the single nucleon sector.  Therefore, a more accurate determination
of the nucleon electric dipole amplitudes is clearly required in order
to make definite predictions for nuclei.  There are many other
processes which can be explored using this technology.  Charged pion
photoproduction is of great interest since in that case the single
nucleon sector is well understood both theoretically and
experimentally.  Also, results ---both scattering lengths and
photoproduction amplitudes---for heavier nuclei like tritium or Helium
would be a particularly novel way to explore the relevance of chiral
symmetry in nuclei.

%%%%%%%%%%%%%%%%%%%%%%%%%%%%%%%%%%%%%%%%%%%%%%%%%%%%%%%%%%%%%%%%%%%%%%%%
\section{Acknowledgements}
We are grateful for valuable
discussions with A. Bernstein, J.L. Friar, U.-G.~Mei{\ss}ner and
M.~Rho. SRB is grateful to the INT (Program INT-95-2 on ``Chiral Dynamics
in Hadrons and Nuclei'') and the nuclear theory group
in Seattle for hospitality while part of this work was completed.
This research was supported in part by the U. S. Department of Energy
(grants DE-FG05-90ER40592 (SRB) and DE-FG06-88ER40427 (UvK)).
%%%%%%%%%%%%%%%%%%%%%%%%%%%%%%%%%%%%%%%%%%%%%%%%%%%%%%%
\appendix
\def\theequation{\Alph{section}.\arabic{equation}}
\setcounter{equation}{0}
\section{Normalization Conventions}

With field normalization convention, the differential cross section
for $\gamma d\rightarrow \pi d $ can  be written as
\begin{equation}
d\sigma\;=\;\frac{(2\pi )^{4}}{u_{\alpha}}\;\delta^{4}(p_{1}+k-p_{2}-q)\;
\frac{{\overline{|{\cal M}|^{2}}}}{2E_{1}\;2E_{2}\;2\omega_{q}\;2\omega_{k}}\;
\frac{d^{3}{\vec p}_{2}}{(2\pi )^{3}}\;\frac{d^{3}{\vec q}}{(2\pi )^{3}},
\end{equation}
where $p_{1}$,$p_{2}$ are the momenta of the initial and final
deuterons, and $q$,$k$ are the momenta of
the outcoming neutral pion and photon, respectively.
$u_{\alpha}$ is the relative velocity of the incident particles,
given by

\begin{equation}
u_{\alpha}\;=\;\frac{p_{1}\cdot k}{E_{1}\;\omega_{k}}.
\end{equation}
In the center of mass frame one finds
\begin{equation}
d\sigma\;=\;\frac{1}{64{\pi }^{2}\;u_{\alpha}}\;
\frac{{\overline{|{\cal M}|^{2}}}}{E_{1}\;E_{2}\;
\omega_{q}\;\omega_{k}}\;\delta (E_{1}+\omega_{k}- E_{2}-\omega_{q}
)\;q^{2}\;dq\;d\Omega .
\end{equation}
The integration over $q$ readily yields

\begin{eqnarray}
d\sigma\; & =\; & \frac{1}{64{\pi}^{2}}\;\frac{E_{1}}{E_{1}+|{\vec k}|}\;
\frac{{\overline{|{\cal M}|^{2}}}}{E_{1}\;E_{2}\;\omega_{q}\;\omega_{k}}\;
q^{2}\;
\frac{E_{2}\;\omega_{q}} {q\;(E_{2}+\omega_{q})}\;d\Omega\nonumber\\ & = &
\frac{1}{64{\pi}^{2}}\;\frac{q}{|{\vec k}|}\;
\frac{{\overline{|{\cal M}|^{2}}}}{(E_{1}+|{\vec k}|)\;
(E_{2}+\omega_{q})}\;d\Omega ,
\end{eqnarray}
where $\omega_{k}=k$, and $E_{1}$, $E_{2}$, and $\omega_{q}$ are the energies
of the deuterons and the pion, respectively.
The slope of the differential cross section at threshold is defined as
\begin{eqnarray}
\frac{|{\vec k}|}{q}\;\frac{d\sigma}{d\Omega}|_{q=0}
& = & \frac{1}{64{\pi }^{2}}\;\frac{{\overline{|{\cal M}|^{2}}}_{q=0}}
{(\sqrt{m_{d}^{2}+
M_{\pi}^{2}}+|\vec{k}|)\;(m_{d}+M_{\pi})}\nonumber\\
&\simeq  & \frac{1}{64{\pi }^{2}}\;
\frac{{\overline{|{\cal M}|^{2}}}_{q=0}}{(m_{d}+M_{\pi})^2}.
\end{eqnarray}

We can express the photoproduction amplitude in terms of rotationally
invariant amplitudes:

\begin{equation}
{\cal M}= i{{\vec J}\cdot {\vec a}}\;{\cal M}_1 +
          i{{\vec J}\cdot {\vec k}}\;{{\vec q}\cdot {\vec a}}\;{\cal M}_2 +
          i{{\vec J}\cdot {\vec q}}\;{{\vec q}\cdot {\vec a}}\;{\cal M}_3 +
          i{{\vec q}\cdot{{\vec k}\times {\vec a}}}{\cal M}_4 ,
\end{equation}
where ${\vec a}\equiv{{\vec\epsilon}}-({{\vec k}\cdot{\vec\epsilon}})
{\vec\epsilon}$.
It is convenient to
define an electric dipole amplitude, $E_d$, such that, in Coulomb gauge,
\begin{equation}
{\cal M}{|_{q=0}}={{\cal M}_d}\equiv {8\pi}(m_{d}+M_{\pi})\;2i({\vec\epsilon}
\cdot{\vec J})\;{E_d},
\end{equation}
where $\;{\vec J}=\frac{1}{2}({\vec \sigma}_{n}+{\vec\sigma}_{p})\;$.
It then follows that

\begin{equation}
\frac{|{\vec k}|}{q}\;\frac{d\sigma}{d\Omega}|_{q=0}=\frac{8}{3}{E_d^2}.
\end{equation}
%%%%%%%%%%%%%%%%%%%%%%%%%%%%%%%%%%%%%%%%%%%%%%%%%%%%%%%%%%%%%%%%%%%%%%%%%%%5
\setcounter{equation}{0}
\section{The S-matrix}

The S-matrix is defined as

\begin{equation}
S_{fi}\;=\;-i(2\pi)^{4}\;\delta^{4}(p_{i}-p_{f})\;(\prod_{i=1}^{n}{\frac{1}
{\sqrt{(2\pi )^{3}2E_{i}}}}  )\;{\cal M},
\end{equation}
where $n$ is the total number of external particles.
For the 3-body process,
$\gamma N N\rightarrow \pi N N$, we find

\begin{equation}
S_{fi}\;=\;-i(2\pi
)^{4}\;\delta^{4}(p_{i}-p_{f})\frac{{\cal M}_{NN}}{(2\pi
)^{9}\;\sqrt{2E_{1}\;2E_{2}\;2E_{1}^{\prime}\;2E_{2}^{\prime}\;2E_{\gamma}
\;2E_{\pi}}},
\end{equation}
whereas the S-matrix for $\;\gamma d\rightarrow \pi
d\;$ is given by

\begin{equation}
S_{fi}\;=\;-i(2\pi
)^{4}\;\delta^{4}(p_{i}-p_{f})\;\frac{{\cal M}_{d}}{(2\pi
)^{6}\;\sqrt{2E_{d}\;2E_{d}^{\prime}\;2E_{\gamma}\;2E_{\pi}}}.
\end{equation}
Therefore,

\begin{equation} {\cal M}_{d}\;=\;\frac{1}{(2\pi
)^{3}}\;\sqrt{\frac{E_{d}\;E_{d}^{\prime}}{4\;E_{1}E_{2}E_{1}^{\prime}
E_{2}^{\prime}}}\;{\cal
M}_{NN}.
\end{equation}
Near threshold,

\begin{equation}
E_{d}\approx\sqrt{m_{d}^{2}+{\vec k}^{2}}\approx {m_d} \;,\;\;\;\;
E_{d}^{\prime}\approx
m_{d}\;,\;\;\;\;E_{1}=E_{2}=E_{1}^{\prime}=E_{2}^{\prime}\approx m ,
\end{equation}
and so we obtain

\begin{equation}
{\cal M}_{d}\;=\;\frac{1}{(2\pi )^{3}\;m}\;{\cal
M}_{NN}.
\end{equation}

%%%%%%%%%%%%%%%%%%%%%%%%%%%%%%%%%%%%%%%%%%%%%%%%%%%%%%%%%%%%%%%%%%%%%%
\setcounter{equation}{0}
\section{Feynman Amplitudes}
First we need the transition operator for $\;\gamma NN\rightarrow\pi NN$.
Figure 3a yields

\begin{eqnarray}
iT_{NN}^{(a)} & = & {\overline{B}}_{v_{1}}\;(\frac{ieg_{A}}{f_{\pi}}\;S^{\mu}_
{v_{1}}\;\epsilon_{a3c}{\tau}_{c}^{1}{\epsilon}_{\mu})\;B_{v_{1}}\;
\frac{i}{{q^{\prime}}^{2}-M_{\pi}^{2}}\;{\overline{B}}_{v_{2}}\;
(\frac{(q^{\prime}+q)_{\nu}\;v_{2}^{\nu}}{4f_{\pi}^{2}}\;
\epsilon_{a3d}\;{\tau}^{2}
_{d})\;B_{v_{2}}\;+\;(1\leftrightarrow 2)\nonumber \\ & = &
-\;\frac{eg_{A}}{4f_{\pi}^{3}}\;{\overline{B}}_{v_{1}}\;
S_{v_{1}}^{\mu}\;B_{v_{1}}\;{\epsilon}_{\mu}\;\frac{(q^{\prime}+q)\cdot
v_{2}}{{q^{\prime}}^{2}-M_{\pi}^{2}}\;{\overline{B}}_{v_{2}}\;B_{v_{2}}\;
\epsilon_{a3c}\epsilon_{a3d}\;{\tau}_{c}^{1}{\tau}_{d}^{2}\;+\;
(1\leftrightarrow 2),
\end{eqnarray}
where $\;{\vec q}^{\prime}={\vec p}-{\vec p}^{\prime}\;$.
We can make use of the relation

\begin{equation}
{\overline{B}}_{v}S^{\mu}_{v}B_{v}
\approx 2m(\frac{1}{2}{\vec \sigma}\cdot {\vec v}\;,
\;\frac{1}{2}{\vec \sigma})
\end{equation}
to obtain

\begin{equation}
iT^{(a)}_{NN}\;=\;-\frac{eg_{A}}{4f_{\pi}^{3}}\;(2m)^{2}\frac{1}{2}\;
{\vec\sigma_{1}}\cdot{\vec\epsilon}\;\frac{2M_{\pi}}
{{\vec{q^{\prime}}}^{2}}\;({\vec \tau}^{1}\cdot {\vec
\tau}^{2}-\tau^{1}_{z}\tau^{2}_{z})\;+\;(1\leftrightarrow 2) .
\end{equation}
Here we approximated $\;q_{0}\approx q_{0}^{\prime}\approx M_{\pi}\;$ and
chose Coulomb gauge ($\;\epsilon_{0}=0\;$ and ${\vec
\epsilon}\cdot{\vec k}=0$).
Using the relation between the transition operator of $\;\gamma NN
\rightarrow\pi NN\;$ and $\;\gamma d\rightarrow\pi d\;$,
obtained in appendix b, we get
\begin{equation}
iT_{d}^{(a)}\;=\;-\frac{2eg_{A}M_{\pi}m}{(2\pi )^{3}f_{\pi}^{3}}\;
\frac{{\vec J}\cdot {\vec\epsilon}}{({\vec p}-{\vec p}^{\prime})^{2}}
\;\;({\vec\tau}^{1} \cdot{\vec\tau}^{2}-\tau^{1}_{z}\tau^{2}_{z}).
\label{seagull}
\end{equation}
The contribution from Figure 3b is

\begin{eqnarray}
iT_{NN}^{(b)} & = &
{\overline{B}}_{v_{1}}\;(-\frac{g_{A}}{f_{\pi}}\;S^{\mu}_
{v_{1}}\;q^{{\prime\prime}}_{\mu}{\tau}_{a}^{1})\;B_{v_{1}}\;
\frac{i}{{q^{{\prime\prime}}}^{2}-M_{\pi}^{2}}\;e\;\epsilon_{a3c}\;
(q^{{\prime\prime}}+q^{\prime})^{\mu}\; \epsilon_{\mu}\nonumber\\ & &
\times \frac{i}{q^{\prime^{2}}-M_{\pi}^{2}}
\;{\overline{B}}_{v_{2}}\;(\frac{1}{4f_{\pi}
^{2}}\;(q^{\prime}+q)_{\nu}\;v_{2}^{\nu}\;\epsilon_{c3b}\;{\tau}^{2}
_{b})\;B_{v_{2}}\;+\;(1\leftrightarrow 2)\nonumber \\ & = &
\;\frac{eg_{A}}{4f_{\pi}^{3}}\;(2m)^{2}\;2M_{\pi}({\frac{1}{2}}
\vec\sigma_{1}\cdot\vec{q}^{{\prime\prime}} )\;
\frac{2\;\vec{q}^{\prime}\cdot\vec\epsilon}
{(\vec{q}^{{\prime\prime}^{2}}+M_{\pi}^{2})\; \vec{q}^{\prime^{2}}}\;
(-)\;\epsilon_{c3a}\epsilon_{c3b}\;{\tau}_{a}^{1}{\tau}_{b}^{2}\;+\;
(1\leftrightarrow 2)\nonumber\\ & = &
-\frac{eg_{A}}{4f_{\pi}^{3}}\;(2m)^{2}\;2M_{\pi}(\vec{J}\cdot
\vec{q}^{{\prime\prime}} ) \frac{2\;\vec{q}^{\prime}\cdot\vec\epsilon}
{(\vec{q}^{{\prime\prime}^{2}}+M_{\pi}^{2})\; \vec{q}^{\prime^{2}}}
\;({\vec\tau}^{1} \cdot{\vec\tau}^{2}-\tau^{1}_{z}\tau^{2}_{z})
\end{eqnarray}
It then follows that
\begin{equation}
iT_{d}^{(b)}\;=\;-\frac{2eg_{A}M_{\pi}m}{(2\pi )^{3}f_{\pi}^{3}}\;
\frac{2(\vec{J}\cdot\vec{q}^{{\prime\prime}})\;
\vec{q}^{\prime}\cdot\vec\epsilon}
{(\vec{q}^{{\prime\prime}^{2}}+M_{\pi}^{2})\; \vec{q}^{\prime^{2}}}
\;({\vec\tau}^{1} \cdot{\vec\tau}^{2}-\tau^{1}_{z}\tau^{2}_{z}),
 \label{pionic}
\end{equation}
where $\vec{q}^{{\prime\prime}} = \vec{p}-\vec{p}^{\prime}-\vec{k}$.
By sandwiching these transition operators between initial and final states,
which include the effect of the nuclear wave functions, we obtain the total
matrix element:
\begin{equation}
{\cal M}_{d}\;=\;<f|T_{d}|i>.
\end{equation}
Finally, we obtain

\begin{eqnarray}
{\cal M}_{d}^{(a)} & =&  i\frac{2eg_{A}M_{\pi}m}{(2\pi )^{3}f_{\pi}^
{3}}<{\Psi_d}|({\vec\tau}^{1}\cdot {\vec\tau}^{2} -
\tau^{1}_{z}\tau^{2}_{z})
\frac{\vec{J}\cdot\vec\epsilon}{\vec{q}^{\prime^{2}}}|{\Psi_d}> \\
{\cal M}_{d}^{(b)} & =&  i\frac{4eg_{A}M_{\pi}m}{(2\pi )^{3}f_{\pi}^
{3}}<{\Psi_d}| ({\vec\tau}^{1}\cdot {\vec\tau}^{2} -
\tau^{1}_{z}\tau^{2}_{z})
\frac{(\vec{J}\cdot\vec{q}^{{\prime\prime}})
\vec{q}^{\prime}\cdot\vec\epsilon}
{(\vec{q}^{{\prime\prime}^{2}}+M_{\pi}^{2}) \vec{q}^{\prime^{2}}}|{\Psi_d}> .
\end{eqnarray}
It is straightforward to check that
$<{\vec\tau}^{1}\cdot {\vec\tau}^{2} -\tau^{1}_{z}\tau^{2}_{z}>=-2$.
%%%%%%%%%%%%%%%%%%%%%%%%%%%%%%%%%%%%%%%%%%%%%%%%%%%%%%%%%%%%%%%%%%%%%%%%%%%%
\setcounter{equation}{0}
\section{Nuclear matrix elements}
\label{int1}

In this appendix we obtain the coordinate space representation of our
matrix elements and evaluate
using the Bonn potential. Consider first the momentum dependent part of
${\cal M}_{D}^{(a)}$:

\begin{eqnarray}
<\frac{{\vec J} \cdot {\vec\epsilon}}{({\vec
p}-\vec{p^{\prime}})^{2}}> & = & \int d^{3}{\vec p}d^{3}{\vec
p}^{\prime}\; {\tilde{\phi}}^{\ast}_{f}({\vec
p}^{\prime})\;\frac{{\vec J}\cdot{\vec\epsilon}}{({\vec p}-{\vec
p}^{\prime})^{2}}\;{\tilde{\phi}}_{i}({\vec p}-\frac{\vec k}{2})\nonumber\\
& = & \frac{1}{4\pi }\;\int d^{3}{\vec r}\,d^{3} {\vec p}\,d^{3}{\vec
p}^{\prime}\;{\tilde{\phi}}^{\ast}_{f}({\vec p}^{\prime})\;({\vec
J}\cdot{\vec\epsilon})\;\frac{e^{-i({\vec p}-{\vec
p}^{\prime})\cdot{\vec r}}}{r}\;{\tilde{\phi}}_{i}({\vec p}-\frac{\vec
k}{2})\nonumber\\ & = & \frac{(2\pi )^{3}}{4\pi }\;\int d^{3}{\vec
r}\;\;\;\;(\frac{1}{(2\pi )^{3/2}}\int d^{3}{\vec
p}^{\prime}\;e^{i{\vec p}^{\prime}\cdot {\vec
r}}\tilde{\phi}^{\ast}_{f}({\vec p}^{\prime}))\nonumber\\ & &
\times\;({\vec J}\cdot{\vec\epsilon})\;\frac{e^{-i\frac{\vec
k}{2}\cdot{\vec r}}}{r}\;\;\;(\frac{1}{(2\pi )^{3/2}}\int d^{3}{\vec
p}\;e^{-i({\vec p}-\frac{\vec k}{2})\cdot {\vec
r}}\tilde{\phi}_{i}({\vec p}-\frac{\vec k}{2}))\nonumber\\ & = &
2\pi^{2}\;\int\;d^{3}{\vec r}\;\phi^{\ast}_{f}({\vec r})\;({\vec
J}\cdot{\vec\epsilon})\;\frac{e^{-i\frac{\vec k}{2}\cdot{\vec
r}}}{r}\;\phi_{i}({\vec r}),
\end{eqnarray}
where $\;\phi\;$ is the spatial part of the deuteron wave function, and the
identity,

\begin{equation}
\frac{1}{{\vec q}^{2}}\;=\;\frac{1}{4\pi}\;\int d^{3}{\vec
r}\;\frac{e^{i{\vec q}\cdot{\vec r}}}{r},
\end{equation}
has been used. The spatial part of the deuteron wave function is

\begin{equation}
\phi({\vec r})\;=\;\frac{1}{\sqrt{4\pi}}\;(\frac{U(r)}{r}+\frac{1}
{\sqrt{8}}S_{12}(\hat{r})\frac{W(r)}{r}),
\end{equation}
where $U(r)$ and $W(r)$ are the S-state and the
D-state of the radial wave function, respectively,
normalized such that $\int_{0}^{\infty}dr(U^{2}+W^{2})=1$.
The spin operator is defined to be

\begin{equation}
S_{12}(\hat{r})\;=\;3(\vec\sigma_{1}\cdot\hat{ r})\;(\vec\sigma_{2}\cdot\hat{
r})-\vec\sigma_{1}\cdot\vec\sigma_{2}.
\end{equation}
Since the S-state wave function has no angular
dependence, the contribution from the S-state is easily evaluated:

\begin{eqnarray}
<\frac{{\vec J} \cdot {\vec\epsilon}}{({\vec p}-{\vec
p}^{\prime})^{2}}>_{S-state} & = & \frac{\pi}{2}\;({\vec
J}\cdot{\vec\epsilon})\;\int d^{3}{\vec
r}\;\frac{U^{2}}{r^{3}}\;e^{-i\frac{\vec k}{2}\cdot{\vec r}}\nonumber\\ & =
& \pi^{2}({\vec J}\cdot{\vec\epsilon})\;\int_{0}^{\infty}
\frac{U^{2}}{r}dr\;\int_{-1}^{1}dx\;e^{-i\frac{kr}{2}x}\nonumber\\
& = & \frac{4\pi^{2}}{k}\;({\vec
J}\cdot{\vec\epsilon})\;\int_{0}^{\infty}
\;\frac{U^{2}}{r^{2}}\;\sin(\frac{kr}{2})\;dr
\end{eqnarray}

Next, we consider the effect of the D-state part of the wave function, which
consists of the cross term
as well as the pure D-component. The cross term is given by
\begin{eqnarray}
<\frac{{\vec J} \cdot {\vec\epsilon}}{({\vec p}-{\vec
p}^{\prime})^{2}}>_{Cross} & = &
\frac{\pi}{2\sqrt{8}}\;\int_{0}^{\infty}dr\frac{UW}{r}\nonumber\\ &  &
\times\;\int d\Omega\; e^{-i\frac{\vec k}{2}\cdot{\vec r}}\;\{{\vec
J}\cdot\vec\epsilon\;S_{12}({\hat{ r}})+S_{12}({\hat{ r}})\;{\vec
J}\cdot\vec\epsilon\} .
\end{eqnarray}
Here we can use
$\;\{\vec\sigma_{i},\;\vec\sigma_{j}\}\;=\;2\delta_{ij}\;$ to show that

\begin{equation}
{\vec J}\cdot\vec\epsilon\;S_{12}({\hat{ r}})+S_{12}({\hat{
r}})\;{\vec J}\cdot\vec\epsilon\;=\;6\;(\vec\epsilon\cdot\hat{
r})({\vec J}\cdot\hat{ r})-2({\vec J}\cdot\vec\epsilon),
\end{equation}
and the cross term becomes

\begin{equation}
<\frac{{\vec J} \cdot {\vec\epsilon}}{({\vec p}-{\vec
p}^{\prime})^{2}}>_{Cross}\; =
\;\frac{\pi}{\sqrt{8}}\;\int_{0}^{\infty}dr\frac{UW}{r}\;\int
d\Omega\; e^{-i\frac{\vec
k}{2}\cdot\vec{r}}\;\{3(\vec\epsilon\cdot\hat{ r})({\vec J}\cdot{\hat
r})-({\vec J}\cdot\vec\epsilon)\}.
\end{equation}

The angular integration yields

\begin{equation}
\int d\Omega\; e^{-i\frac{\vec k}{2}\cdot\vec{
r}}\;(\vec\epsilon\cdot\hat{ r})({\vec J}\cdot{\hat r})\;=\;({\vec
J}\cdot\vec\epsilon)\;\frac{4\pi}{a^{3}}(\sin{a}-a\cos{a}),
\end{equation}
where $a=kr/2$. Finally we have

\begin{equation}
<\frac{{\vec J} \cdot {\vec\epsilon}}{({\vec p}-{\vec
p}^{\prime})^{2}}>_{Cross} = \frac{8{\pi}^{2}\sqrt{2}}{k^{3}}\;({\vec
J}\cdot{\vec\epsilon})\;\int_{0}^{\infty}dr
\frac{UW}{r^{4}}\;\{3\sin{\frac{kr}{2}}-3\frac{kr}{2}\cos{\frac{kr}{2}}-
(\frac{kr}{2})^{2}\sin{\frac{kr}{2}}\}.
\end{equation}\

The pure D-state contribution can be evaluated in similar fashion:

\begin{eqnarray}
<\frac{{\vec J} \cdot {\vec\epsilon}}{({\vec p}-{\vec
p}^{\prime})^{2}}>_{D-state} & = & \frac{\pi}{16}\int
dr\;\frac{W^{2}}{r}\;\int d\Omega\;e^{-i\frac{\vec k}{2}\cdot {\vec
r}}\;S_{12}^{\ast}\;({\vec J}\cdot \vec\epsilon )\;S_{12}\nonumber\\ & = &
\frac{\pi}{16}\int dr\;\frac{W^{2}}{r}\;\int d\Omega\;e^{-i\frac{\vec
k}{2}\cdot {\vec r}}\;6(2(\vec\epsilon\cdot\hat{ r})({\vec
J}\cdot\vec\epsilon)-({\vec J}\cdot\vec\epsilon))\nonumber\\ & = &
\frac{3\pi^{2}}{k^{3}}({\vec J}\cdot\vec{\epsilon})\;\int
dr\frac{W^{2}}{r^{4}}\;(8\sin\frac{kr}{2}-4\cos
\frac{kr}{2}-\sin\frac{kr}{2}).
\end{eqnarray}
Evaluating the integrals with the
Bonn potential yields
\begin{eqnarray}
<\frac{{\vec J} \cdot {\vec\epsilon}} {({\vec p}-{\vec
p}^{\prime})^{2}}>_{S-state} & = &({\vec
J}\cdot{\vec\epsilon})\;7.616\; fm^{-1}\nonumber\\
<\frac{{\vec J} \cdot {\vec\epsilon}} {({\vec p}-{\vec
p}^{\prime})^{2}}>_{Cross} & = & ({\vec
J}\cdot{\vec\epsilon})\;0.051\; fm^{-1}\nonumber\\
<\frac{{\vec J} \cdot {\vec\epsilon}} {({\vec p}-{\vec
p}^{\prime})^{2}}>_{D-state} & = & ({\vec
J}\cdot{\vec\epsilon})\;(-0.090) \; fm^{-1}\nonumber
\end{eqnarray}
The contributions from the D-state wave function are
clearly negligible relative to the S-state contributions.

Finally, consider the momentum dependent part of ${\cal M}_{D}^{(b)}$.
Here we make use of the integral parametrization:

\begin{eqnarray}
\frac{1}{\{(\vec{q}^{\prime}
-\vec{k})^{2}+M_{\pi}^{2}\}\;{\vec{q}}^{\prime^{2}}} & = &
\int^{1}_{0}\;\frac{dz}{[\{(\vec{q}^{\prime}
-\vec{k})^{2}+M_{\pi}^{2}\}z\;+\;(1-z){\vec{q}}^{\prime^{2}}]^{2}}
\nonumber\\ & = & \int^{1}_{0}\;\frac{dz}{({\vec{\it l}}^{2} \;+
\;m^{{\prime}^{2}})^{2}}\nonumber ,
\end{eqnarray}
where ${\vec{\it l}}\equiv\vec{q}^{\prime} -z\vec{k}$,
$m^{{\prime}^{2}}\equiv M_{\pi}^{2}z(2-z)$, and
$\vec{q}^{\prime}=\vec{p}-\vec{p}^{\prime}$ .
We can then write

\begin{equation}
\frac{\vec{J}\cdot(\vec{q}^{\prime} -\vec{k})\;\;\vec{\epsilon}
\cdot\vec{q}^{\prime}}{\{(\vec{q}^{\prime} -\vec{k})^{2}+M_{\pi}^{2}\}\;
{\vec{q}}^{\prime^{2}}} \;=\;
\int^{1}_{0}\;dz\;\frac{(\vec{J}\cdot{\vec{\it l}})\;
(\vec{\epsilon}\cdot {\vec{\it l}})\;+\;(z-1)\;(\vec{J}\cdot\vec{k})\;
(\vec{\epsilon}\cdot {\vec{\it l}})}{({\vec{\it l}}^{2} \;+
\;m^{{\prime}^{2}})^{2}}\nonumber .
\end{equation}
The second term vanishes in the pure S-state.
The Fourier transform of the first term is

\begin{equation}
\frac{(\vec{J}\cdot{\vec{\it l}})\;
(\vec{\epsilon}\cdot {\vec{\it l}})}{({\vec{\it l}}^{2} \;+
\;m^{{\prime}^{2}})^{2}}\;=\; \frac{1}{8\pi}\;\int\;d^{3}\vec{r}\;
e^{-i{\vec{\it l}}\cdot \vec{r}}\;\{\frac{\vec{J}\cdot
\vec{\epsilon}}{r} -
(\vec{J}\cdot\hat{r})\;(\vec{\epsilon}\cdot\hat{r})
\;(\frac{1}{r}+m^{\prime})\} \;e^{-m^{\prime}r}\nonumber ,
\end{equation}
which, in turn, yields the matrix element
\begin{equation}
<\;\frac{(\vec{J}\cdot{\vec{\it l}})\;
(\vec{\epsilon}\cdot {\vec{\it l}})}{(\vec{\it l}^{2} \;+
\;m^{\prime^{2}})^{2}}\;>\;=\;\frac{1}{8\pi}\;\int\;d^{3}\vec{p}\;d^{3}
{\vec{p}}^{\prime}\;d^{3}\vec{r}\;\phi_{f}({\vec{p}}^{\prime})\;
e^{-i(\vec{p}-{\vec{p}^{\prime}}-z\vec{k})\cdot \vec{r}} \;f(\vec{r})\;
e^{-m^{\prime} r}\;\phi_{i}(\vec{p}-\frac{\vec{k}}{2})\nonumber ,
\end{equation}
where $\;f(\vec{r})\;\equiv\;\frac{\vec{J}\cdot \vec{\epsilon}}{r} \;-\;
(\vec{J}\cdot \hat{r})\;(\vec{\epsilon}\cdot\hat{r})
\;(\frac{1}{r}+m^{\prime})$ .
Considering the pure S-state effect, we obtain
\begin{eqnarray}
\lefteqn{<\frac{\vec{J}\cdot(\vec{q}^{\prime} -\vec{k})\;
\;\vec{\epsilon}\cdot\vec{q}^{\prime}}{\{(\vec{q}^{\prime}
-\vec{k})^{2}+m^{2}\}\;{\vec{q}}^{\prime^{2}}}{>_{S-state}}
\;=\;\int_{0}^{1}dz\;<\;\frac{(\vec{J}\cdot\vec{\it
l})\;(\vec{\epsilon}\cdot\vec{\it l})}{(\vec{\it l}^{2}\;+
\;m^{\prime^{2}})^{2}}{>_{S-state}}}\nonumber\\
& = &
{\pi}^{2}\int^{1}_{0}dz\int\;d^{3}{\vec{r}}\;\phi_{f}(\vec{r})
\; (\vec{J}\cdot\vec{\epsilon})\;\frac{e^{-m^{\prime}r}}{r}\;
e^{i(z-\frac{1}{2})\vec{k}\cdot\vec{r}}\; \phi_{i}(\vec{r})\nonumber\\
& & -\pi^{2}\int_{0}^{1}dz\int\;d^{3}\vec{r}\;
\phi_{f}(\vec{r})\; (\vec{J}\cdot\hat{r})\;(\vec{\epsilon}\cdot
\hat{r}) \;(\frac{1}{r}+m^{\prime})\;e^{-m^{\prime}r}\;
e^{i(z-\frac{1}{2})\vec{k}\cdot\vec{r}}\; \phi_{i}(\vec{r})\nonumber\\
& = & \pi^{2}(\vec{J}\cdot\vec{\epsilon})\int_{0}^{1}dz
\int_{0}^{\infty}dr\;e^{-m^{\prime}r} \;\frac{U^{2}}{r}\;
\frac{\sin{[(z-\frac{1}{2})kr]}}{(z-{\frac{1}{2}})kr}\nonumber \\
& & -\pi^{2}(\vec{J}\cdot\vec{\epsilon})\;\int_{0}^{1}dz
\int_{0}^{\infty}dr\;e^{-m^{\prime}r}\;U^{2}\;(\frac{1}{r}+m^{\prime})\;
\{\frac{\sin(z-{\frac{1}{2}} )kr}{[(z-{\frac{1}{2}})kr]^{3}}\;-\;
\frac{\cos(z-{\frac{1}{2}})kr}{[(z-{\frac{1}{2}})kr]^{2}} \}.\nonumber\\
\end{eqnarray}

Finally we obtain

\begin{equation}
<\;\frac{(\vec{J}\cdot\vec{q}^{{\prime\prime}})\;
\vec{q}^{\prime}\cdot\vec\epsilon}
{(\vec{q}^{{\prime\prime}^{2}}+M_{\pi}^{2})\;
\vec{q}^{\prime^{2}}}\;{>_{S-state}}
 = <\vec{J}\cdot\vec\epsilon >\;0.747 \;f^{-1}_{m}
\end{equation}
where we have used  $k_{th}=0.685f_{m}^{-1}$ (mass of $\pi_{0}$).

\begin{table}[t]
\centering
\begin{tabular}{||c|c|c|c|c||}     \hline
&${E_{0+}^{{\pi ^0}p}}$  &${E_{0+}^{{\pi ^0}n}}$  &$E^{ss}_{d}$  &$E_d$ \\
\hline
$\vartheta (q^3)\;{\rm incomplete}$  &-2.24  &0.5  &-1.33 &-3.97 \\ \hline
$\vartheta (q^3)$  & 0.96  &3.7  & 3.6 &0.94 \\ \hline
experiment       &-2.0$\pm$0.2 (?)  &?    &-     &-3.74$\pm$0.25 \\ \hline
\end{tabular}
\caption[blah]{The importance of the single-scattering contribution:
The ${\chi PT}$ predictions at
$\vartheta (q^3)$ without the triangle graphs ---serving as a
phenomenological estimate--- lead to agreement
with the experimental value of $E_d$\cite{arg}.}
\end{table}

\begin{figure}[t]
   \vspace{0.5cm}
   \epsfysize=7cm
   \centerline{\epsffile{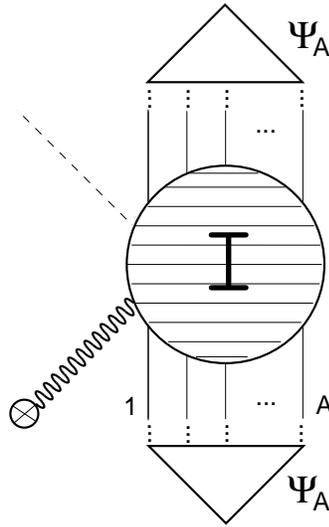}}
   \centerline{\parbox{11cm}{\caption{\label{fig}
The anatomy of a matrix element. The $\Psi _A$'s correspond to the
nuclear wave functions. The blob is the sum of all $A$-nucleon
irreducible graphs.
  }}}
\end{figure}

\begin{figure}[t]
   \vspace{0.5cm}
   \epsfysize=12cm
   \centerline{\epsffile{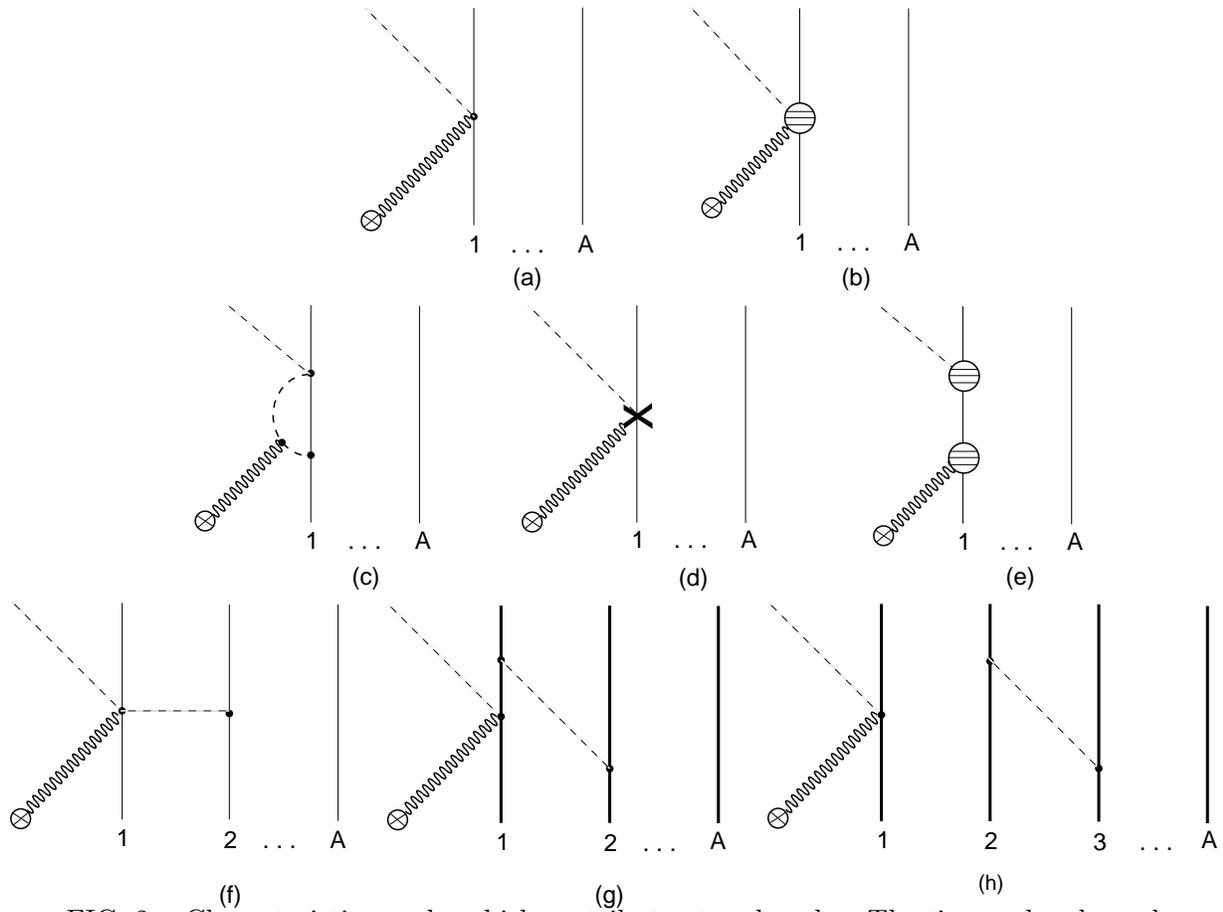}}
   \centerline{\parbox{11cm}{\caption{\label{fig}
Characteristic graphs which contribute at each order. The time-ordered
graphs are distinguished by bold nucleon lines.
  }}}
\end{figure}

\begin{figure}[t]
   \vspace{0.5cm}
   \epsfysize=4cm
   \centerline{\epsffile{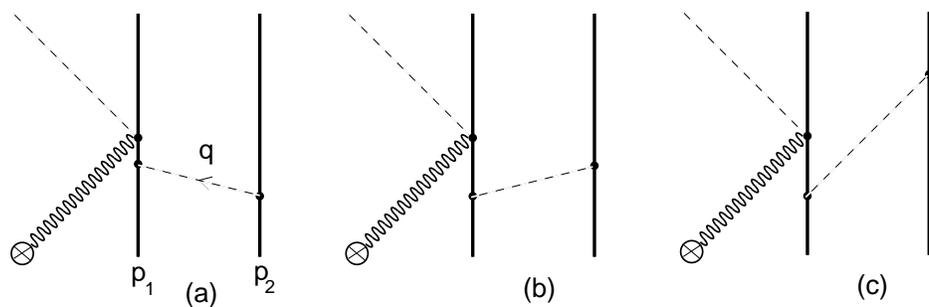}}
   \centerline{\parbox{11cm}{\caption{\label{fig}
The energy dependence of the impulse approximation graphs (a) and (b)
approximately cancels the time-ordered 3-body diagram (c).
  }}}
\end{figure}

\begin{figure}[t]
   \vspace{0.5cm}
   \epsfysize=7cm
   \centerline{\epsffile{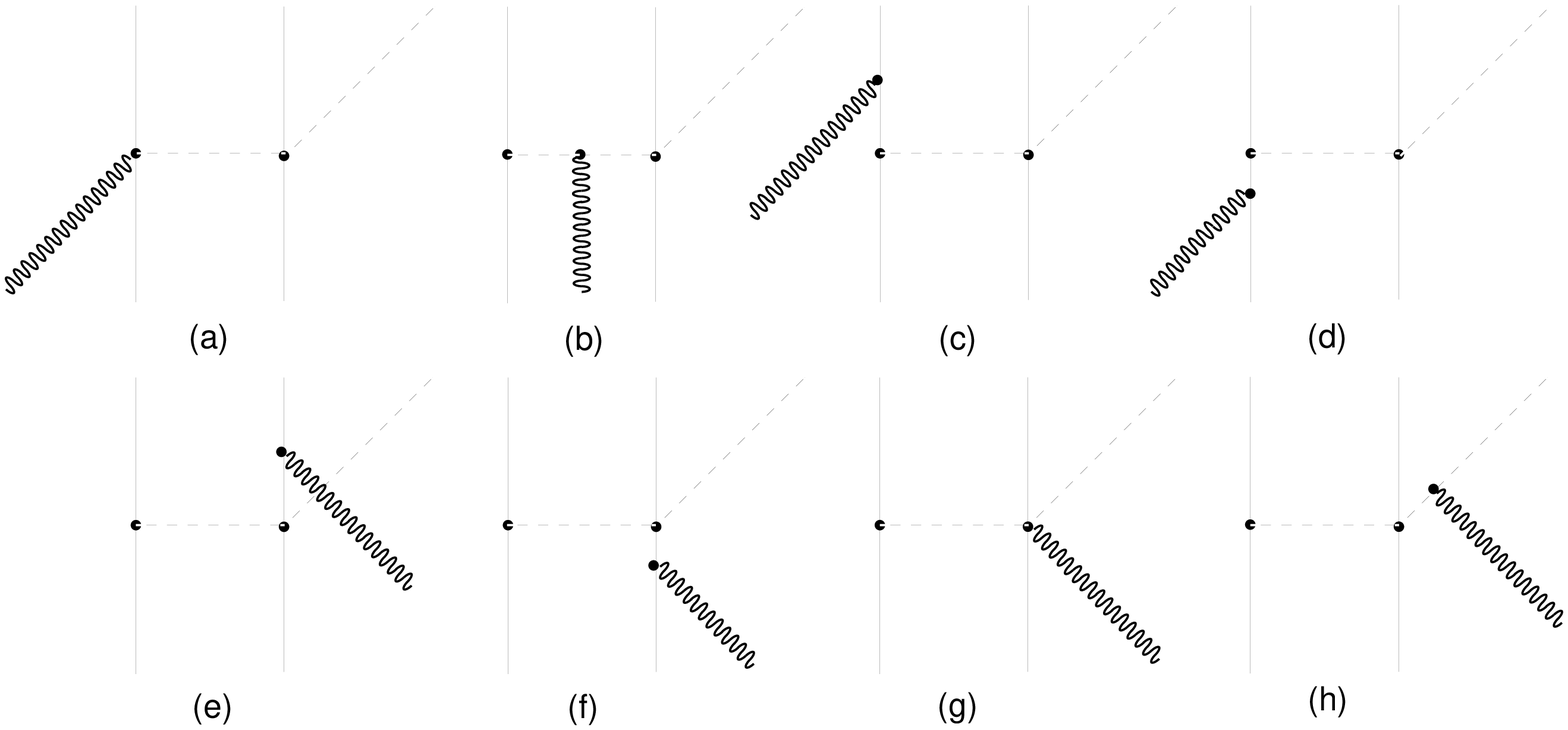}}
   \centerline{\parbox{11cm}{\caption{\label{fig}
Gauge invariant subset of 3-body interactions. In Coulomb gauge, only graphs
(a) and (b) contribute to neutral pion photoproduction.
  }}}
\end{figure}


\newpage
\begin{thebibliography}{99}
\bibitem{swnp1} S.~Weinberg, Phys. Lett. B {\bf 251}, 288 (1990).
\bibitem{swnp2} S.~Weinberg, Nucl. Phys. {\bf B363}, 3 (1991).
\bibitem{swnp3} S.~Weinberg, Phys. Lett. B {\bf 295}, 114 (1992).
\bibitem{ubi1} C.~Ord\'{o}\~{n}ez and U.~van Kolck, Phys. Lett. B {\bf 291},
               459 (1992).
               C.~Ord\'{o}\~{n}ez, L.~Ray and U.~van Kolck, Phys. Rev. Lett.
               {\bf 72}, 1982 (1994).
\bibitem{ubi2} U. van Kolck, Phys. Rev. C {\bf 49}, 2932 (1994).
\bibitem{ubi3} U. van Kolck, U. of Washington preprint DOE/ER/40427-13-N94,
                             in preparation.
\bibitem{rho2}  M.~Rho, Phys. Rev. Lett. {\bf 66}, 1275 (1991).
\bibitem{rho} T.-S.~Park, D.-P.~Min and M.~Rho, Phys. Rep. {\bf 233}, 341
              (1993).
\bibitem{physca} S.~Weinberg, Physica {\bf 96A}, 327 (1979).
\bibitem{ulf}  Ulf-G.~Mei{\ss}ner, Rep. Prog. Phys. {\bf 56}, 903 (1993).
\bibitem{ulf2}  V.~Bernard, N.~Kaiser, and Ulf-G.~Mei{\ss}ner, Strasbourg
                preprint CRN-95-3, hep-ph/9501384.
\bibitem{bonn} R.~Machleidt, K.~Holinde, and Ch.~Elster, Phys. Rep. {\bf 149},
               1 (1987).
\bibitem{hg} H.~Georgi, Phys. Lett. B {\bf 240}, 447 (1990).
\bibitem{jm} E.~Jenkins and A.V.~Manohar, Phys. Lett. B {\bf 255}, 558 (1991).
\bibitem{ktb} S.S.~Kamalov, L.~Tiator and C.~Benhold, Phys. Rev. Lett.
              {\bf 75}, 1288 (1995).
\bibitem{ber1} V.~Bernard, J.~Gasser, N.~Kaiser, and Ulf-G.~Mei{\ss}ner, Phys.
               Lett. B {\bf 268}, 291 (1991).
\bibitem{tia} D.~Drechsel and L.~Tiator, J. Phys. G  {\bf 18}, 449 (1992).
\bibitem{vain} P.~de Baenst, Nucl. Phys. {\bf B24}, 633 (1970);
               I.A.~Vainshtein and V.I.~Zakharov,  Sov. J. Nucl. Phys.
               {\bf 12}, 333 (1971); Nucl. Phys. {\bf B36}, 589 (1972).
\bibitem{nkf}  H.W.L.~Naus, J.H.~Koch and J.L.~Friar, Phys. Rev. C {\bf 41},
               2852 (1990).
\bibitem{mainz} R.~Beck et al, Phys. Rev. Lett. {\bf 65}, 1841 (1990).
\bibitem{saclay} E.~Mazzucato et al, Phys. Rev. Lett. {\bf 57}, 3144 (1986).
\bibitem{bh} A.M.~Bernstein and B.R.~Holstein,  Comm. Nucl. Part. Phys.
             {\bf 20}, 197 (1991).
\bibitem{ber1a} V.~Bernard, N.~Kaiser, and Ulf-G.~Mei{\ss}ner, Nucl. Phys.
                {\bf B383}, 442 (1992).
\bibitem{ber2} V.~Bernard, N.~Kaiser, J.~Kambor, and Ulf-G.~Mei{\ss}ner,
               Nucl. Phys. {\bf B388}, 315 (1992).
%\bibitem{gasser} See, for example, J.~Gasser and Ulf.G.~Mei{\ss}ner,
                  Nucl. Phys. {\bf B357}, 90 (1991).
\bibitem{skf}  S.~Scherer, J.H.~Koch and J.L.~Friar, Nucl. Phys. {\bf A552},
               515 (1993).
\bibitem{ber3} V.~Bernard, N.~Kaiser, and Ulf-G.~Mei{\ss}ner, Strasbourg
               preprint CRN-94-62, hep-ph/9411287.
\bibitem{arg} P.~Argan et al, Phys. Rev. Lett. {\bf 41}, 629  (1978);
                              Phys. Rev. C {\bf 24}, 300  (1981).
\bibitem{koch} J.H.~Koch and R.M.~Woloshyn, Phys. Rev. C {\bf 16}, 1968 (1977).
\bibitem{laget} P.~Bosted and J.~Laget, Nucl. Phys. {\bf A296}, 413 (1978).
\bibitem{faldt} G.~F{\" a}ldt, Phys. Scrip. {\bf 22}, 5 (1980).
\end{thebibliography}
\end{document}